\documentclass[12pt]{article}

\newcommand{\x}{arXiv:}
\newcommand{\m}{\mathrm}
\newcommand{\be}{\begin{equation}}
\newcommand{\ee}{\end{equation}}
\newcommand{\ba}{\begin{eqnarray}}
\newcommand{\ea}{\end{eqnarray}}

\usepackage{graphicx}
\usepackage{amssymb}
\usepackage{amsmath}
\usepackage[T1]{fontenc} 
\usepackage[ansinew]{inputenc} 
\usepackage[nosort]{cite}
\newcommand{\inbar}{\vrule height1.57ex width.4pt depth0pt}
\newcommand{\SW}{\relax{\hbox{$\ \inbar\kern-.285em{\rm S}$}}}

\begin{document}
\thispagestyle{empty}
\begin{center}

\null \vskip-1truecm \vskip2truecm

{\Large{\bf \textsf{Holographic Dual of The Weak Gravity Conjecture}}}

{\large{\bf \textsf{}}}

{\large{\bf \textsf{}}}

\vskip1truecm

{\large \textsf{Brett McInnes}}

\vskip1truecm

\textsf{\\  National
  University of Singapore}

\textsf{email: matmcinn@nus.edu.sg}\\

\end{center}
\vskip1truecm \centerline{\textsf{ABSTRACT}} \baselineskip=15pt
\medskip

The much-discussed \emph{Weak Gravity Conjecture} is interesting and important in both the asymptotically flat and the asymptotically AdS contexts. In the latter case, it is natural to ask what conditions it (and the closely related Cosmic Censorship principle) imposes, via gauge-gravity duality, on the boundary field theory. We find that these conditions take the form of lower bounds on the number of colours in this theory: that is, the WGC and Censorship might (depending on the actual sizes of the bounds) enforce the familiar holographic injunction that this number should be ``large''. We explicitly estimate lower bounds on this number in the case of the application of holography to the quark-gluon plasma produced in heavy ion collisions. We find that classical Censorship alone prohibits realistically small values for the number of colours, but that the WGC offers hope of resolving this problem.

\newpage

\addtocounter{section}{1}
\section* {\large{\textsf{1. The Instability of (Near-)Extremal Black Holes}}}
Increasing the charge on a black hole\footnote{All black holes in this work, except where explicitly noted, have strictly zero angular momentum.} of fixed mass reduces its entropy; the temperature also drops, and it turns out that a finite amount of charge, dependent on the mass of the black hole, apparently suffices to reduce the temperature to zero. This shows that the black hole, if it is indeed to continue to be a black hole, cannot accept more than this amount of charge. A black hole carrying this maximal charge, such that its temperature is zero, is said to be \emph{extremal}.

Thus we arrive at a remarkable conclusion: the charge and mass of a black hole cannot be entirely independent of each other. Given the mass, we can put an upper bound on the charge of non-spinning, that is, Reissner-Nordstr\"{o}m (henceforth, RN) black holes. In the strictly classical context, this is the statement of \emph{Cosmic Censorship} in this case. Notice that black hole thermodynamics \cite{kn:wall} allows us to discuss Censorship without passing judgement as to the undesirability of naked singularities.

One way to understand Censorship is to study the stability of black holes as they near extremality, that is, exactly zero temperature: perhaps extremal black holes cannot be stable. This is strongly supported on general thermodynamic grounds, since (in the RN case) an extremal black hole has a small but non-zero entropy with a non-universal value. This would not normally be considered acceptable. Notice that this argument can also be used in the \emph{near}-extremal case, since a system with an extremely low temperature but with entropy bounded away from zero is also unacceptable from this perspective.

More specifically, the statement that extremal black holes must always be able to decay (to a near-extremal black hole, which might itself be unstable) by emitting particles is the \emph{Weak Gravity Conjecture} \cite{kn:motl,kn:palti,kn:rude} (henceforth, WGC)\footnote{The fact that an extremal black hole can decay in this manner, despite having zero temperature, is explained clearly in \cite{kn:ong}.}. If this conjecture is correct --$\,$ and there is mounting evidence that it is, see for example \cite{kn:rem,kn:shiu,kn:goon} --$\,$ then a large charge on a black hole will trigger the emission of charged particles, to the effect that extremal and (sufficiently) near-extremal black holes cannot be stable in string theory.

Classical Censorship conditions have to be corrected when higher-dimension operators are taken into account. Thinking about ``quantum Censorship'' in terms of the WGC has the advantage that it allows us to predict at least the general form that these corrections should take: if we require that they should not obstruct the decay of extremal black holes, then we can sketch the shape that the modified locus of points describing extremal black holes should take in the mass-charge plane \cite{kn:kats}. This is very useful. For example, we learn from it that the WGC-modified definition of extremality is such that, given the mass, \emph{larger values of the charge are permitted} than would be allowed in the classical case\footnote{It is argued in \cite{kn:cano1,kn:cano2} that this may not necessarily be the case for all black holes when $\alpha^{\prime}$ corrections are fully taken into account. We will assume that it is correct for the black holes we consider here, but these complications will have to be considered in a complete theory.}. The WGC can be regarded as a more precise, quantum-corrected, version of classical Censorship: we can speak of a ``WGC/Censorship'' condition.

The WGC/Censorship condition can be formulated on an anti-de Sitter background\footnote{See for example \cite{kn:qing,kn:mig,kn:crem,kn:agar}, and, most directly relevant here, \cite{kn:naka1,kn:naka2}. For the asymptotically \emph{de Sitter} case in four dimensions, see for example \cite{kn:qing,kn:anton}.}. Doing so leads immediately to a natural question: \emph{what is the holographic dual \cite{kn:casa,kn:nat,kn:bag} of this condition?} WGC/Censorship\footnote{It is important to note that, in the case of four-dimensional, asymptotically flat and de Sitter black holes, doubts regarding the validity of Censorship have recently been expressed  \cite{kn:horava,kn:bambi,kn:card,kn:vitor,kn:hans,kn:fel,kn:roy,kn:casals}. But in the five-dimensional, asymptotically AdS$_5$ case most relevant to holography, there is strong evidence \cite{kn:suvrat,kn:bala,kn:sean} (see also \cite{kn:wang}) that Censorship \emph{does} hold universally. Furthermore, the idea that the WGC ``enforces'' Censorship is particularly convincing in the asymptotically AdS situation \cite{kn:weak,kn:horsant}.}, as a constraint on the mass and charge, must translate to a constraint on the quantities \emph{dual} to the mass and the charge. The implications of such a constraint, \emph{which may have no obvious basis in terms of boundary physics}, are clearly of great interest.

We find that the holographic dual of WGC/Censorship for AdS$_5$-RN black holes implies a lower bound on $N_{\textsf{c}}$, the number of colours describing the boundary field theory. As is well known, the number of colours is normally taken to be ``large'' in applications of holography \cite{kn:99}; but this is usually presented as a matter of expediency: it suppresses the most difficult quantum-gravitational corrections. Here we are arguing that, \emph{if} our lower bound is ``large'', then ``large'' values of $N_{\textsf{c}}$ are \emph{enforced} by WGC/Censorship\footnote{The cases in which $N_{\textsf{c}}$ is not \emph{arbitrarily} large are quite distinct, in several interesting ways, from the more familiar cases in which it is. See for example \cite{kn:tor1,kn:oka,kn:tor2,kn:tor3}.}.

We study this in a specific example, namely the well-known application of the duality to the \emph{quark-gluon plasma}. We find that the classical Censorship condition for AdS$_5$-RN black holes, evaluated using actual data from phenomenological models and from experiment, forces the number of colours to be considerably larger than the value in QCD. To put it another way: any attempt to construct a holographic version of QCD will have to deal with the tendency of the bulk black hole to violate Censorship as the number of colours approaches a realistic value.

On the other hand, we are able to argue that the WGC significantly reduces the bound imposed by classical Censorship on $N_{\textsf{c}}$. Thus, it may be that the WGC will play a decisive role in future attempts to construct realistic holographic models.

Classical Censorship for AdS$_5$-RN black holes takes an unfamiliar form, since the constraint imposed on the charge and the mass is not linear, as it is in the asymptotically flat case. This immediately leads to a potential problem: even if the WGC is satisfied at one value of the mass, it does not obviously follow that it is satisfied for all masses. Our first task is to confirm that no problems of this sort arise. Having done this, we can proceed to deduce an explicit bound on $N_{\textsf{c}}$ in terms of the other parameters of the boundary field theory.

\addtocounter{section}{1}
\section* {\large{\textsf{2. Classical Censorship for AdS$_5$-Reissner-Nordstr\"om Black Holes}}}
The AdS$_5$-Reissner-Nordstr\"om black hole metric is
\begin{flalign}\label{A}
g(\m{RNAdS}_5)\;=\;&-\,\left({r^2\over L^2}\,+\,1\,-\,{2M\over r^2}\,+\,{Q^2\over r^4}\right)\m{d}t^2\,+{\m{d}r^2\over {r^2\over L^2}\,+\,1\,-\,{2M\over r^2}\,+\,{Q^2\over r^4}}\\ \notag \,\,\,\,&\,+\,r^2\left(\m{d}\theta^2 \,+\, \sin^2\theta\,\m{d}\phi^2\,+\,\cos^2\theta\,\m{d}\psi^2\right).
\end{flalign}
Here $L$ is the asymptotic curvature scale, $r$ and $t$ are as usual, and we take it that the $r = $ constant sections are three-spheres described by Hopf coordinates. The geometric parameters $M$ and $Q$ both have units of squared length. In this Section and the next (only) we will use five-dimensional Planck units, defined so that $M$ and $Q$ are the physical mass and charge\footnote{Planck units are defined here by setting the five-dimensional Planck length equal to $(3 \pi /4)^{1/3}$, and the five-dimensional Coulomb constant equal to $4 \pi^2$. In ``particle theory'' units, the physical mass and charge of the black hole are fixed multiples of $M$ and $Q$: see below, equations (\ref{J}).}.

If Censorship holds, the black hole has an outer horizon located at $r = r_{\textsf{H}}$, which satisfies
\begin{equation}\label{B}
{r_{\textsf{H}}^6\over L^2}\,+\,r_{\textsf{H}}^4\,-\,2M\,r_{\textsf{H}}^2\,+\,Q^2\;=\;0.
\end{equation}
This is a cubic equation in $r_{\textsf{H}}^2$.

The behaviour of the roots of the general cubic $ax^3 + bx^2 + cx + d$ is governed by the \emph{discriminant} \cite{kn:disc},
\begin{equation}\label{C}
\Delta \;\equiv\; 18abcd - 4b^3d + b^2c^2 - 4ac^3 - 27a^2d^2.
\end{equation}
With the pattern of signs in the case at hand, one finds that the left side of (\ref{B}), regarded as a cubic, always has one negative real root. It has two distinct positive real roots if and only if $\Delta$ is strictly positive, and it has one positive real root if and only if $\Delta = 0$.

Using this, we find that the condition for Censorship to hold here is
\begin{equation}\label{D}
{27\over 4 L^4}Q^4\,+\,\left(1 + {9M\over L^2}\right)Q^2 \,-\, M^2\left(1 + {8M\over L^2}\right)\;\leq\;0.
\end{equation}
This is not useful as it stands: we need to separate $Q$ from $M$. To do this, we regard the expression on the left as a quadratic in $Q^2$. The discriminant of this quadratic can be expressed, after some manipulation, as $\left(1 + 6M/L^2\right)^3$, and so (\ref{D}) can be put into the form
\begin{equation}\label{E}
{|Q|\over L^2}\;\leq \;{\sqrt{2} \over 3\sqrt{3}}\left[\left(1 + {6M\over L^2}\right)^{3/2} -\; \left(1 + {9M\over L^2}\right)\right]^{1/2}.
\end{equation}
This is the Censorship condition for these black holes.

If the right side of (\ref{E}) is expanded as a power series in $M/L^2$, one finds, to lowest order in $M/L^2$, that $L$ drops out, and this inequality becomes simply $|Q| \leq M$; which is indeed the statement of Censorship for an asymptotically flat five-dimensional Reissner-Nordstr\"{o}m black hole (obtained by taking the limit $L \rightarrow \infty$). That is, Censorship takes the familiar linear form in the asymptotically flat case, but \emph{not} in the asymptotically AdS case.

Before proceeding, we note for later use that there is a simple formula \cite{kn:disc} for the repeated roots of a cubic when $\Delta = 0$, that is, when (\ref{E}) is saturated. Using that formula and combining it with (\ref{E}), one can find an expression for $r_{\textsf{H}}^{\textsf{ext}}$, the value of $r_{\textsf{H}}$ in the extremal case:
\begin{equation}\label{EA}
{\left(r_{\textsf{H}}^{\textsf{ext}}\right)^2\over L^2}\;=\;{{1\over 3}\left[\left(1 + {6M\over L^2}\right)^{3/2} -\; \left(1 + {9M\over L^2}\right)\right] + {M\over L^2}\over 1 + {6M\over L^2}}.
\end{equation}
Notice that this does not vanish, in sharp contrast to the behaviour of the event horizons of AdS$_5$-Kerr black holes as extremality is approached \cite{kn:96}.

\begin{figure}[!h]
\centering
\includegraphics[width=0.7\textwidth]{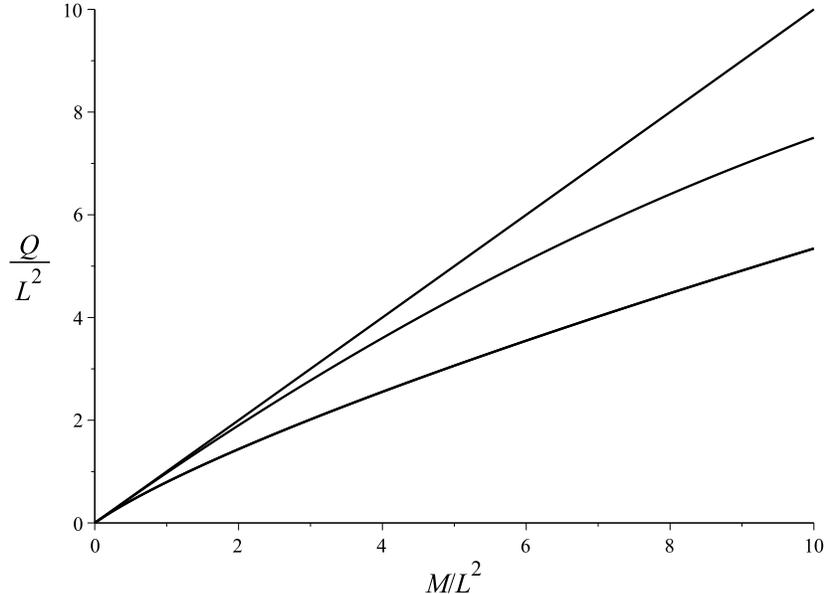}
\caption{Uppermost curve: extremality for asymptotically flat five-dimensional Reissner-Nordstr\"{o}m black holes. Lowermost curve: extremality for asymptotically AdS$_5$ Reissner-Nordstr\"{o}m black holes. Middle curve: possible locus of points representing quantum-corrected extremal asymptotically AdS$_5$ Reissner-Nordstr\"{o}m black holes. Event horizons are spherical in all cases. The middle curve is asymptotic to the lowermost curve as $M/L^2$ increases, but comes close to it only at extremely large $M/L^2$.}
\end{figure}

The function on the right in (\ref{E}) is shown as the lowermost curve in Figure 1: all parameter values corresponding to points below the curve, and only those, satisfy classical Censorship. (We take it here, and henceforth, that $Q$ is positive.) The function is unbounded, but its slope tends monotonically  to zero at large $M/L^2$; we will discuss this further, below. For comparison, the straight line corresponding to the asymptotically flat case is also drawn, as the uppermost curve. (In this case, $L$ is some arbitrary quantity with units of length.) The two curves are close only for small values of $M/L^2$ and $Q/L^2$. (The meaning of the middle curve will be explained below.)

\addtocounter{section}{1}
\section* {\large{\textsf{3. Instability of Extremal AdS$_5$-Reissner-Nordstr\"om Black Holes}}}
Consider an extremal AdS$_5$-RN black hole, corresponding to a point on the lowermost curve in Figure 1, with mass and charge $(M, Q)$ in five-dimensional Planck units. Assume that the black hole can emit a particle of mass $m$ and positive charge $q$, after which its parameters become $(M + \delta M, Q + \delta Q)$, where $\delta M$ and $\delta Q$ are of course negative. Then by the triangle inequality (which is valid on the tangent spaces of any Lorentzian manifold), we have
\begin{equation}\label{F}
M + \delta M + m \leq M,
\end{equation}
as well as $q = - \delta Q$, and so we have
\begin{equation}\label{G}
{q \over m}\,\geq \,{\delta Q \over \delta M}.
\end{equation}
(For convenience, we now set $M^* = M/L^2, Q^* = Q/L^2$.) On the other hand, the requirement that the translation $\left(M^*, Q^*\right) \rightarrow \left(M^* + \delta M^*, Q^* + \delta Q^*\right)$ should move the point from being on the lowermost curve in Figure 1 to being below it --$\,$ that is, the requirement that classical Censorship should continue to hold --$\,$ means that we must have
\begin{equation}\label{H}
{\delta Q \over \delta M}\, > \, {\m{d}Q^* \over \m{d}M^*}\left(M^*, Q^*\right).
\end{equation}
Thus we have finally
\begin{equation}\label{I}
{q \over m}\, > \, {\m{d}Q^* \over \m{d}M^*}\left(M^*, Q^*\right).
\end{equation}

In the asymptotically flat case, the right side of (\ref{I}) is identically equal to unity, and so we obtain the well-known consequence: if the extremal black hole is to decay by radiating charged particles, these particles must satisfy $q/m > 1$, a result sometimes described, figuratively, as meaning that ``gravity is the weakest force''. It is conjectured that black holes too must be able to satisfy this inequality, meaning that the definition of extremality is modified accordingly, as we discussed earlier. This would be represented by a curve in Figure 1 (not drawn) which must lie \emph{above} the straight line $Q = M$, though it must be asymptotic to the latter at large $M$: see \cite{kn:kats}. Thus we see very simply that the WGC allows larger values of $Q$, for a given $M$, than does classical Censorship. (Note that, in all cases, the graphs for small $M/L^2$ have only a formal meaning, as they correspond to ``black holes'' with sub-Planckian masses.)

Now we turn to the asymptotically AdS$_5$ case. For these black holes, when $M^*$ and $Q^*$ are close to zero, the right side of (\ref{I}) is approximately equal to unity (as is to be expected, since $L \rightarrow \infty$ is the asymptotically flat limit). For all larger values, however, it is strictly \emph{smaller} than unity, as is evident from Figure 1. If the WGC is satisfied in the case of arbitrarily large $L$ --$\,$ that is, if there exist particles with $q/m > 1$ --$\,$ then \emph{it is satisfied in all cases}, that is, for all values of all parameters, including $L$.

We stress that this was result was not inevitable: the lowermost graph in Figure 1 might well have had one of several different possible shapes incompatible with the WGC for some values of the parameters. We can see this with the aid of a very interesting example, as follows.

As is well known \cite{kn:lemmo}, in the asymptotically AdS case there exist black holes with exotic event horizon topologies. In particular, the event horizon can be a flat, cubic three-torus, T$^3$. The metric of a toral AdS$_5$-RN black hole has the form
\begin{flalign}\label{IA}
g(\m{RNAdS^{T^3}}_5)\;=\;&-\,\left({r^2\over L^2}\,-\,{2M\over r^2}\,+\,{Q^2\over r^4}\right)\m{d}t^2\,+{\m{d}r^2\over {r^2\over L^2}\,-\,{2M\over r^2}\,+\,{Q^2\over r^4}}\\ \notag \,\,\,\,&\,+\,r^2\left(\m{d}\theta^2 \,+\,\m{d}\phi^2\,+\,\m{d}\psi^2\right),
\end{flalign}
where now the angular coordinates are polar coordinates on a unit circle, with a periodicity $2\pi K$, where $K$ is dimensionless and can take any positive value. Thus the $r = $ constant sections are modelled on a torus of dimensionless ``volume'' $V = 8\pi^3K^3$. The parameters $M$ and $Q$ are not the physical mass and charge: the physical mass is $\mathcal{M} = M/V$, and the physical charge is $\mathcal{Q} = Q/V$. (We have compactified the event horizon so as to avoid irrelevant complications with infinite parameter values; see below.)

Equation (\ref{B}) becomes now
\begin{equation}\label{IB}
{r_{\textsf{H}}^6\over L^2}\,-\,2M\,r_{\textsf{H}}^2\,+\,Q^2\;=\;0,
\end{equation}
and the requirement that the discriminant of this cubic in $r_{\textsf{H}}^2$ should be non-negative is simply (with positive $Q$)
\begin{equation}\label{IC}
{Q\over L^2}\;\leq \;\left({32\over 27}\right)^{1/4}\,\left({M\over L^2}\right)^{3/4}.
\end{equation}
This is the classical Censorship condition in this case. The graph of the function on the right is shown as Figure 2, with the straight line representing Censorship for the asymptotically flat case shown for comparison. (There are no asymptotically flat black holes with toral event horizons; this explains why the two curves need not have the same slope at the origin.)
\begin{figure}[!h]
\centering
\includegraphics[width=0.7\textwidth]{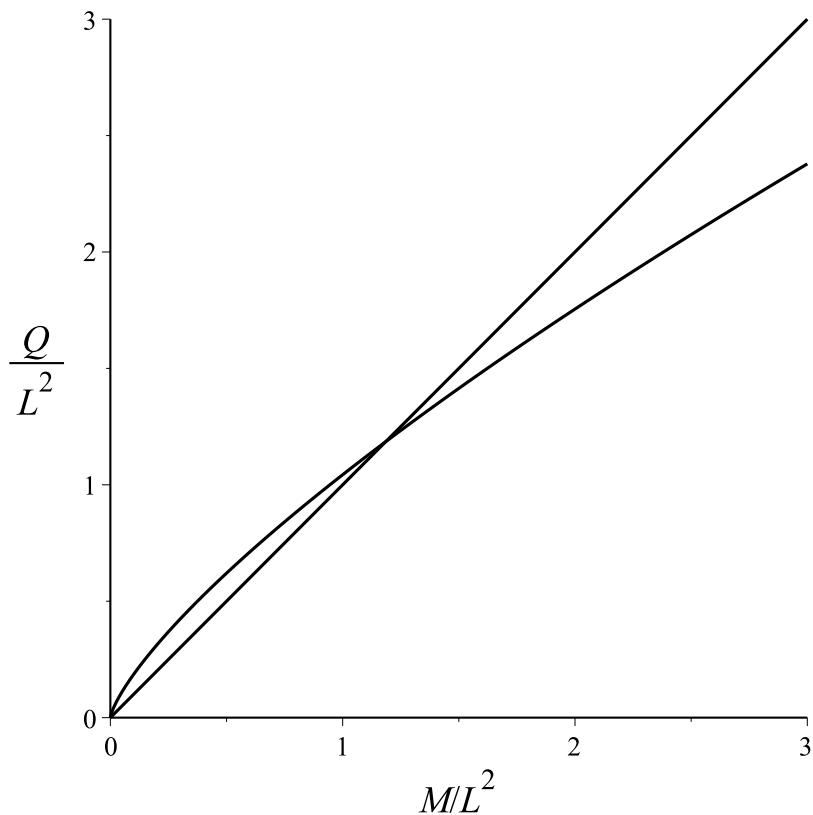}
\caption{Classical Censorship for asymptotically AdS$_5$ Reissner-Nordstr\"{o}m black holes (with toral event horizons). (Classical Censorship for asymptotically flat five-dimensional Reissner-Nordstr\"{o}m black holes, $Q/L^2 = M/L^2$, shown for comparison.)}
\end{figure}
The inequality (\ref{I}) now becomes
\begin{equation}\label{ID}
{q \over m}\, > \, \left({3 V L^2\over 8 \mathcal{M}}\right)^{1/4}.
\end{equation}
The right side can be made arbitrarily large by selecting a suitably small value for $\mathcal{M}$, or a suitably large one for $V$. Therefore, the inequality (\ref{I}) cannot be satisfied for all parameter values in this case unless [a] there are infinitely many varieties of stable particles with $q/m > 1$, a situation which the WGC itself is designed to avoid \cite{kn:motl}, or [b] there is some upper bound on $VL^2/\mathcal{M}$. A lower bound on the mass is quite possible, but an upper bound on $V$ is hard to accept. In particular, it would mean that the event horizons of AdS black holes with flat event horizons \emph{must} be compactified, and that the compactification scale is bounded by a quantity determined by the maximal value of $q/m$.

In short, the shape of the graph in this case means that these black holes are not compatible with the WGC, at least not for all parameter values. The fact that the WGC does work for all parameter values, in the case of interest to us here, is not trivial.

Conversely, one could claim that the WGC \emph{predicts} that the graph must have the shape shown as the lowermost curve in Figure 1, even in the classical case. When quantum corrections are taken into account \cite{kn:kats}, as discussed above, one expects to find a similar graph, again with slope everywhere smaller than unity; it would fit between\footnote{We are assuming that the WGC always permits larger charges for a given black hole mass than does classical Censorship. This is certainly the case for asymptotically flat black holes \cite{kn:kats}; it must persist in the asymptotically AdS case, because, for any given $M = \hat{M}$, no matter how large, it must be possible to make the two cases coincide for all $M \leq \hat{M}$ to any desired accuracy by taking $L$ to be sufficiently large.} the upper and lower graphs in Figure 1, but would be asymptotic to the lower graph (because the WGC should agree with classical Censorship for black holes with sufficiently large masses). An example of such a shape is shown as the middle curve in Figure 1: that is, \emph{points on that curve correspond to black holes which are extremal after quantum corrections have been taken into account}. (We stress that this curve is chosen to illustrate the point; it is not the result of a detailed calculation.)

Thus the WGC gives us a clear, if not altogether precise, picture of the form taken by quantum Censorship in this case. Notice however that the WGC and classical Censorship are very similar for extremely large values of $M/L^2$. If we wish to use the WGC to improve on classical Censorship in some application, then we need to demonstrate that $M/L^2$ takes some value which, in a sense we need to clarify, is not excessively large.

In an AdS$_5$ background, we can use the gauge-gravity duality to apply these findings to the dual system on the boundary. We now turn to the details of that.

\addtocounter{section}{1}
\section* {\large{\textsf{4. The Holographic Interpretation of WGC/Censorship}}}
In this section we set up a holographic dual of the Censorship condition, the inequality (\ref{E}). We work at first with classical Censorship, returning to the WGC at the end of this Section.

Henceforth, we drop five-dimensional Planck units and use instead natural units, with either the electron volt or the femtometre (1 fm $\approx (197.3$ MeV)$^{-1}$) as the base, with the exception that we do not normalize the five-dimensional Coulomb constant\footnote{Our convention will be that electric charge in five dimensions, as in four, is dimensionless. This has the important consequence that $k_5$ is not dimensionless, as its counterpart is in four dimensions: instead, it has units of length.}, $k_5$.

In these units, we have to distinguish the parameters $M$ and $Q$, which describe the geometry of the black hole, from the latter's actual physical mass\footnote{In particular, if we assert that a certain black hole has a ``large'' mass, this should refer to $\mathcal{M}$, not $M$.} $\mathcal{M}$ and physical charge $\mathcal{Q}$. With $M$ and $Q$ appearing in the manner they do in equation (\ref{A}), the relations between these quantities are given by
\begin{equation}\label{J}
\mathcal{M}\;=\;{3\pi M\over 4\ell^3_{\textsf{P}}},\;\;\;\;\;\mathcal{Q}\;=\;{\sqrt{3}\pi^{3/2} Q\over \ell^{3/2}_{\textsf{P}}\sqrt{k_5}},
\end{equation}
where $\ell_{\textsf{P}}$ is the AdS$_5$ Planck length. It is straightforward to express the Censorship condition (\ref{E}) in terms of $\mathcal{M}$ and $\mathcal{Q}$.

The black hole entropy per unit physical mass, which we denote by $\mathcal{S}$, is
\begin{equation}\label{K}
\mathcal{S}\;=\;{2\pi r_{\textsf{H}}^3\over 3 M};
\end{equation}
here we have used the first member of (\ref{J}). Note that $r_{\textsf{H}}$ can be regarded as a function of $M$, $Q$, and $L$ by solving equation (\ref{B}).

The asymptotic value of the electromagnetic potential form (computed, as usual, by requiring that the potential form should be regular at the event horizon) is
\begin{equation}\label{L}
\mathbb{A}_{\infty}\;=\;-\,{\sqrt{3}Q \sqrt{k_5}\over 4\sqrt{\pi} \ell^{3/2}_{\textsf{P}}r_{\textsf{H}}^2}\m{d}t;
\end{equation}
here we have used the second member of (\ref{J}).

Holography \cite{kn:casa,kn:nat,kn:bag} identifies $\mathcal{S}$ with $s/\varepsilon$, where $s$ is the entropy density, and $\varepsilon$ is the energy density of the boundary matter. The holographic interpretation of $k_5$ is given by noting that $1/\left(4 \pi^2k_5\right)$ is precisely the energy corresponding to a pair of unit charges in the bulk, separated by $\delta r = k_5$. We therefore regard $1/\left(4 \pi^2k_5\right)$ as the characteristic energy scale, $E_{\textsf{F}}$, of the dual field theory. (In the application to the quark-gluon plasma, this would be the QCD scale, $\approx $ 220 MeV.)

The disproportion between matter and antimatter in the boundary field theory is measured \cite{kn:bran} by the \emph{baryonic chemical potential}, $\mu_{\textsf{B}}$, which is dual \cite{kn:clifford} to a fixed multiple of the coefficient of d$t$ in $\mathbb{A}_{\infty}$; we have, eliminating $k_5$ in favour of $E_{\textsf{F}}$,
\begin{equation}\label{M}
\mu_{\textsf{B}}\;=\;{3 \sqrt{3} Q \over 8 \pi^{3/2} \ell_{\textsf{P}}^{3/2}\sqrt{E_{\textsf{F}}}r_{\textsf{H}}^2}.
\end{equation}

Finally, the ``holographic dictionary'', which states that
\begin{equation}\label{N}
\ell_{\textsf{P}}^3\; =\; {\pi L^3 \over 2 N_{\textsf{c}}^2},
\end{equation}
allows us to replace the bulk quantity $\ell_{\textsf{P}}$ with the boundary quantity $N_{\textsf{c}}$, the number of colours characterising the boundary field theory:
\begin{equation}\label{O}
\mu_{\textsf{B}}\;=\;{3 \sqrt{3} Q N_{\textsf{c}}\over 4 \sqrt{2}  \pi^{2} \sqrt{E_{\textsf{F}}} L^{3/2} r_{\textsf{H}}^2}.
\end{equation}

Now if, as above, we regard $r_{\textsf{H}}$ as a function of $M$, $Q$, and $L$, then (\ref{K}) and (\ref{O}) give us a pair of equations which, in principle, can be solved to express $Q$ and $M$ as functions of $\mathcal{S} = s/\varepsilon$, $\mu_{\textsf{B}}$, $L$, $N_{\textsf{c}}$, and $E_{\textsf{F}}$ . \emph{Then the Censorship condition (\ref{E}) predicts a novel relationship between these quantities}. This is the dual interpretation of the Censorship condition for these black holes.

Unfortunately, the expressions resulting from directly proceeding in this way are extremely complicated, because one has to find an explicit solution of a cubic, and then solve a pair of non-linear simultaneous equations; when the solutions are substituted into (\ref{E}), the result is not surveyable.

We can circumvent this problem by substituting (\ref{K}) (with $\mathcal{S}$ interpreted as $s/\varepsilon$)  and (\ref{O}) directly into equation (\ref{B}), obtaining
\begin{equation}\label{P}
{r_{\textsf{H}}^2 \over L^2}\;-\;{4 \pi r_{\textsf{H}}\over 3 (s/\varepsilon)}\;+\;{32 \pi^4 E_{\textsf{F}} \mu_{\textsf{B}}^2 L^3\over 27 N_{\textsf{c}}^2} \;+\; 1 \;=\;0.
\end{equation}
Regarding $s/\varepsilon$, $\mu_{\textsf{B}}$, and the other parameters as being fixed, we see that this is a quadratic in $r_{\textsf{H}}$, so the Censorship condition is just
\begin{equation}\label{Q}
- {4 \pi^2 L^2\over 9 (s/\varepsilon)^2} \;+\;1\;+\;{32 \pi^4 E_{\textsf{F}} \mu_{\textsf{B}}^2 L^3\over 27 N_{\textsf{c}}^2}\;\leq \;0.
\end{equation}
Thus we obtain finally the \emph{Censorship bound} on the number of colours in the boundary field theory:
\begin{equation}\label{R}
N_{\textsf{c}}\;\geq \; {4 \sqrt{2} \pi^2 \sqrt{E_{\textsf{F}}L^3} \mu_{\textsf{B}} \over 3 \sqrt{3} \sqrt{{4 \pi^2 L^2 \over 9 (s/\varepsilon)^2} - 1}}.
\end{equation}

Of course, $\mu_{\textsf{B}}$ and $s/\varepsilon$ will vary from one case to another, depending on how the strongly coupled matter they describe is prepared. However, they both have maximal possible values on the parameter domain in which a description by such matter is possible. The right side of (\ref{R}) is an increasing function of these parameters, and so, if we evaluate it at the maximal values of $\mu_{\textsf{B}}$ and $s/\varepsilon$, then we will obtain a universal (for a given kind of strongly coupled matter) lower bound on $N_{\textsf{c}}$.

In short: \emph{Censorship puts a lower bound on the number of colours in the boundary field theory}\footnote{The form of the denominator on the right shows that it also puts a lower bound on $L$: evidently we must have $L > 3 s/(2 \pi \varepsilon)$. In practice, the values of $L$ we discuss below are far larger than this.}, in terms of the boundary data $s/\varepsilon$, $\mu_{\textsf{B}}$, $L$, and $E_{\textsf{F}}$.

In a particular application, one might be able to estimate all of the quantities on the right side of (\ref{R}), and so obtain an explicit bound, and we will see how this works in the next Section. In the general case, we can regard all of these quantities as ``known'' properties of the boundary system, with the possible exception of $L$.

However, we can obtain a weaker bound of this kind even if $L$ is completely unknown, by simply noting that the right side of (\ref{R}), regarded as a function of $L$ with all other parameters fixed, has a positive global minimum. This minimum can be straightforwardly computed, and the result is
\begin{equation}\label{S}
N_{\textsf{c}}\;\geq \;3^{3/4} \sqrt{2\pi}\, (s/\varepsilon)^{3/2} E_{\textsf{F}}^{1/2} \mu_{\textsf{B}}.
\end{equation}
Again, one uses this by inserting the maximal values of $\mu_{\textsf{B}}$ and $s/\varepsilon$ in a given situation.

As we will see, (\ref{S}) is in practice much weaker than Censorship itself, (\ref{R}); but it has the virtue of showing that $N_{\textsf{c}}$ is bounded below, whatever the value of $L$ may be.

The conclusion is simple, yet fundamental: when we assume, in using holography, that $N_{\textsf{c}}$ must be ``large'', this may be no mere matter of avoiding some mathematical complications. It may be a direct consequence of Censorship. In the case where the bulk geometry approximates that of an AdS$_5$-RN black hole, \emph{if} the bounds in (\ref{R}) or (\ref{S}) are large, then large $N_{\textsf{c}}$ is \emph{required} by the Censorship condition for these black holes. We will see below that, in a concrete application, the lower bounds on $N_{\textsf{c}}$ are indeed predicted to be large (though not ``extremely large'').

This is not necessarily a particularly welcome conclusion, since the actual value of $N_{\textsf{c}}$ in QCD is of course quite small: the hope has been that choosing large $N_{\textsf{c}}$ is (in principle, if not in practice) optional, so that holography might ultimately be extended towards (more) realistic values when technical issues can be overcome. Censorship seemingly rules this out.

However, we have been using here classical Censorship. As we saw, the WGC permits larger values of $Q$, for given $M$, than classical Censorship, at least for values of $M/L^2$ which are not extremely large. Without a much more detailed account of the exact way in which the WGC modifies the strictures imposed by classical Censorship, it is not possible to specify exactly what happens to our bounds if the former replaces the latter. However, in qualitative terms, the result is rather clear.

The WGC permits the existence of extremal black holes that would not be allowed by classical Censorship: that is, by the inequality (\ref{R}). It follows that the WGC must replace the right side of (\ref{R}) by some quantity which is \emph{smaller}. In other words, the WGC weakens the inequality (\ref{R}) (and therefore also (\ref{S}), which is derived from it). \emph{One might even hope that the WGC could push the bounds down to the point where realistic (that is, of the order of 3) values of $N_{\textsf{c}}$ become possible}. The WGC, then, could possibly play a central role in constructing more nearly realistic holographic models of strongly coupled field theories.

We have stressed repeatedly that this would only be possible if $M/L^2$ is (in some sense to be explained) not extremely large, since otherwise the WGC essentially coincides with classical Censorship. To assess all this, we need a concrete example, such that the bounds in (\ref{R}) and (\ref{S}) can be evaluated explicitly. We now turn to that.

\addtocounter{section}{1}
\section* {\large{\textsf{5. Example: The Quark-Gluon Plasma at the RHIC}}}
A well-known application of the gauge-gravity duality \cite{kn:casa,kn:nat,kn:bag} is to the study of the Quark-Gluon Plasma produced in collisions \cite{kn:loiz}, at a variety of definite impact energies, at the RHIC facility\footnote{The QGP is also studied at the LHC, but we will not consider those collisions here, because the baryonic chemical potentials of the corresponding plasmas are completely negligible, and we will need numerical values of this parameter in the sequel.} \cite{kn:STAR}. Observational and phenomenological data are available for the entropy and energy densities $s$ and $\varepsilon$ \cite{kn:sahoo} and for the baryonic chemical potential $\mu_{\textsf{B}}$ \cite{kn:sahoo,kn:STARchem} at each impact energy; the characteristic energy scale here is that of QCD, $\approx$ 220 MeV. Thus we can insert explicit values for $s/\varepsilon$, $\mu_{\textsf{B}}$, and $E_{\textsf{F}}$ into the right sides of the inequalities (\ref{R}) and (\ref{S}).

The values drawn from \cite{kn:sahoo,kn:STARchem} are shown in the table. The first column is the impact energy per pair, in GeV, for collisions observed in different runs. The second column is $s/\varepsilon$, in MeV$^{-1}$, and the third column is the baryonic chemical potential, in MeV. (For full details, including error estimates, see the references; it would be pointless to belabour such matters here.)

\begin{center}
\begin{tabular}{|c|c|c|}
  \hline
$\sqrt{s_{\textsf{NN}}}$&$s/\varepsilon$&$\mu_{\textsf{B}}$ \\
\hline
$11.5$ & 0.006822  &400 \\
$14.5$ &  0.006807  &300 \\
$19.6$ &  0.006779  &220 \\
$27$  &  0.006775 & 160 \\
$39$  &  0.006702  & 120 \\
$62.4$  &  0.006684  &80 \\
$200$  &  0.006515  &30 \\
\hline
\end{tabular}
\end{center}

Now the right sides of (\ref{R}) and (\ref{S}) are increasing functions of both $\mu_{\textsf{B}}$ and $s/\varepsilon$, so, as explained earlier, we are interested in the largest values of these quantities. It is clear from the table that these correspond to the lowest impact energies. On the other hand, at very low impact energies, it is questionable whether the QGP actually forms, so we cannot use arbitrarily low energies. It is suggested in \cite{kn:sahoo} that the threshold is somewhere between $\sqrt{s_{\textsf{NN}}} = 11.5$ and $\sqrt{s_{\textsf{NN}}} = 19.6$ GeV; for simplicity we therefore take the values at $\sqrt{s_{\textsf{NN}}} = 14.5$ GeV. Inserting those values into (\ref{S}), we find
\begin{equation}\label{T}
N_{\textsf{c}}\;\geq \; \approx 14.3.
\end{equation}

We stress again that this is a very conservative bound; the true value on the right is undoubtedly larger. The point is that we do not need to know $L$ here.

If we are willing to estimate $L$, then we can use the exact (classical) Censorship bound, (\ref{R}). There are various ways in which one might attempt to do so: here is one.

In this discussion, we have concentrated exclusively on central collisions. However, peripheral collisions are also of great interest and have recently yielded very interesting data, again from the RHIC experiments \cite{kn:STARcoll,kn:STARcoll2,kn:zuo}. In this case, the holographic model of the QGP involves AdS$_5$-Kerr black holes \cite{kn:schalm,kn:93,kn:irina,kn:garb}. Apart from a small set of parameter values \cite{kn:100}, which do not occur here, it turns out that such black holes always satisfy a simple inequality:
\begin{equation}\label{U}
\mathcal{A}\;<\;L,
\end{equation}
where $\mathcal{A}$ is the ratio of the angular momentum of the black hole to its mass. This quantity is dual to the ratio of the angular momentum density of the ``vortical plasma'' to its energy density, which can be estimated in phenomenological models \cite{kn:jiang}. Thus (\ref{U}) gives us an explicit lower bound on $L$. Let us assume that the values of $L$ permitted in this case are the same as in the collisions we have been discussing (the reasoning being that one case can be continuously deformed into the other, by reducing the centrality to approximately zero).

One finds \cite{kn:93,kn:95} that, for collisions at impact energy $\sqrt{s_{\textsf{NN}}} = 200$ GeV per pair and centrality $17\%$, $\mathcal{A}$ attains its largest value for these experiments, $\mathcal{A} \approx 77$ femtometres; so $L$ must be at least this large\footnote{Since $L$ sets the curvature scale of the boundary spatial sections, the fact that $L$ is so large relative to all of the other scales in the problem means that these sections are essentially flat, which is of course highly desirable in itself.}. With all of the other parameters fixed, the right side of (\ref{R}) is a monotonically increasing function of $L$ for all values of $L$ in this range, so we are now in a position to put a lower bound on the right side of (\ref{R}).

Again, we use the data corresponding to $\sqrt{s_{\textsf{NN}}} = 14.5$ GeV for $s/\varepsilon$ and $\mu_{\textsf{B}}$, and we find now that
\begin{equation}\label{V}
N_{\textsf{c}}\;\geq \; \approx 97.2.
\end{equation}
This is an explicit form of the statement that $N_{\textsf{c}}$ must be ``large''.

Any attempt to push the number of colours down, from values like this, towards realistic values, will apparently be blocked by classical Censorship. However, we argued above that the WGC might solve or alleviate this problem. To assess that, we need to see whether the black holes used in this application correspond in some way to the domain in which the WGC differs significantly from classical Censorship.

Let us begin with the AdS$_5$-RN black hole describing the plasmas generated in central RHIC collisions at impact energy $\sqrt{s_{\textsf{NN}}} = 14.5$ GeV as above. Let us fix $M$ and imagine increasing $Q$ towards its extremal value; we think of this as defining a trajectory in Figure 1, a trajectory that takes us, from a starting point below all three curves, towards the lowest curve, where we cross over into the parameter domain permitted by the WGC but forbidden by classical Censorship. The objective is to argue that the crossing point is in some intermediate range, where the ``WGC curve'' does indeed differ substantially from the ``classical Censorship curve''.

Proceeding along this trajectory, we see that $r_{\textsf{H}}$ decreases monotonically (towards $r_{\textsf{H}}^{\textsf{ext}}$, see equation (\ref{EA}) above). The Hawking temperature of this black hole can be expressed as
\begin{equation}\label{VA}
4\pi T_{\textsf{H}}\; =\; {6 r_{\textsf{H}} \over L^2}\,+\, {4\over r_{\textsf{H}}}\,-\,{4 M\over r_{\textsf{H}}^3},
\end{equation}
which decreases monotonically as $r_{\textsf{H}}$ decreases, eventually reaching zero at classical extremality.

On the other hand, from equation (\ref{O}) we see that $\mu_{\textsf{B}}$ steadily \emph{increases} from its already large initial value, around 300 MeV. The trajectory we are studying in the $(M, Q)$ plane is therefore mapped holographically to a trajectory in the quark matter phase diagram, which begins in the region describing the quark-gluon plasma, and ends in the region corresponding to low-temperature but very high-density forms of strongly coupled matter, such as may exist in the cores of certain neutron stars \cite{kn:annala}. The baryonic chemical potential for such matter is indeed large, but it is not vastly larger than 300 MeV \cite{kn:baym}. We take this as a sign that our starting point is not actually very remote from extremality, in the sense that the physical parameters do not change very much as we proceed along the trajectory described above.

Our objective is to compute, or at least bound, $M/L^2$ for the black holes along this trajectory. Substituting equation (\ref{EA}) into equation (\ref{K}), we obtain the extremal value, $\mathcal{S}^{\textsf{ext}}$, of the specific entropy:
\begin{equation}\label{W}
{\mathcal{S}^{\textsf{ext}}\over L}\;=\;{2\pi \left[{{1\over 3}\left[\left(1 + {6M\over L^2}\right)^{3/2} -\; \left(1 + {9M\over L^2}\right)\right] + {M\over L^2}\over 1 + {6M\over L^2}}\right]^{3\over 2}\over 3 M/L^2}.
\end{equation}
The graph of this function is shown as Figure 3. Notice that there is a maximum at $M/L^2 = 4/3$.
\begin{figure}[!h]
\centering
\includegraphics[width=0.7\textwidth]{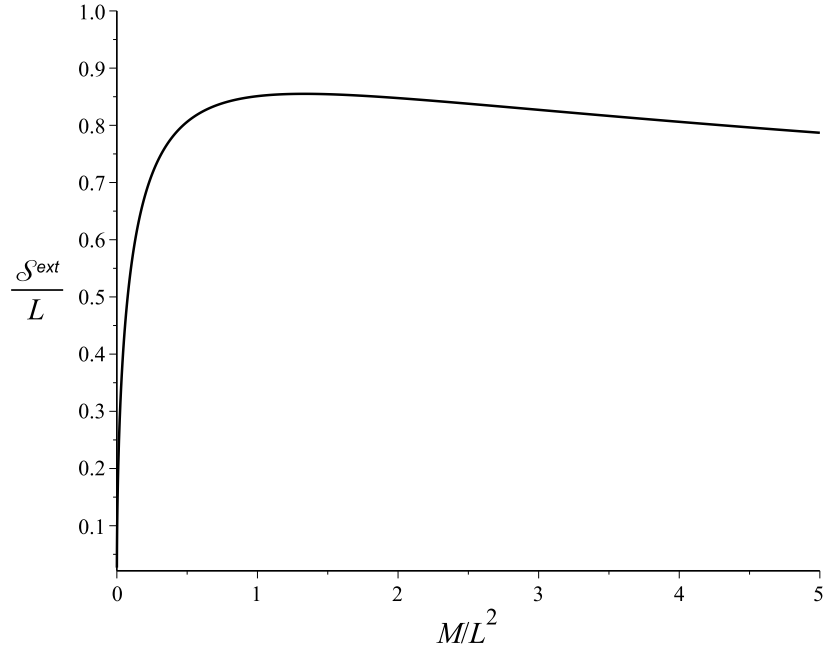}
\caption{The extremal value of the specific entropy of an AdS$_5$-RN black hole, as a function of $M/L^2$.}
\end{figure}

If we insert a value for the left side, then (\ref{W}) can easily\footnote{When, that is, it can be solved at all: as is clear from the Figure, this equation has no real solutions when $\mathcal{S}/L$ is larger than approximately $0.855$. Thus, with $L = 77$ fm, we find that the specific entropy of the boundary fluid cannot be larger than about $0.334$ MeV$^{-1}$. In practice, this is far above realistic values for the QGP, as one can see from the table above; but it is interesting, as illustrating the pervasive power of Censorship, that the latter constrains the entropy of the black hole and therefore the entropy density of the field theory.} be solved numerically for $M/L^2$.

There are always two solutions. One of these is necessarily smaller than $4/3$, the other larger. It turns out that the smaller solution corresponds to ``black holes'' with masses smaller than the five-dimensional Planck mass, so we will not consider these further.

To investigate the meaning of the larger solution, note that the specific entropy at extremality cannot be larger than its initial value (before we began to increase $Q$). In the case of the collisions at $\sqrt{s_{\textsf{NN}}} = 14.5$ GeV, with $s/\varepsilon \approx 0.006807$ MeV$^{-1}$, this means (with $L >  77$ fm) that we have an upper bound on the left side of (\ref{W}), and consequently the larger solution of the equation gives us a lower bound on $M/L^2$. This lower bound is surprisingly large compared to the input: we have
\begin{equation}\label{X}
{M\over L^2}\; > \; \approx 6.16 \times 10^7.
\end{equation}
To understand this, let us use it to compute the ratio of the physical mass of the black hole to the (five-dimensional) Planck mass, $m_{\textsf{P}} \equiv 1/\ell_{\textsf{P}} $. Combining equations (\ref{J}) and (\ref{N}), we have
\begin{equation}\label{Y}
{\mathcal{M}\over m_{\textsf{P}}}\; =\;3\, \left({\pi \over 16}\right)^{{1\over 3}}\,N_{\textsf{c}}^{4/3}\,{M\over L^2}.
\end{equation}
Using (\ref{V}) and (\ref{X}), we now have
\begin{equation}\label{Y}
{\mathcal{M}\over m_{\textsf{P}}}\;>\; \approx 4.80 \times 10^{10}.
\end{equation}
Again, we stress that we had no reason to expect that these simple equations would lead to such a large output.

A black hole with a mass of (say) 100 billion Planck masses is perhaps massive enough to be semi-classical, but not so massive as to be indistinguishable from a fully macroscopic object. We tentatively conclude that we are now precisely in the domain where the WGC differs significantly from classical Censorship. That is, when our constant-$M$ trajectory crosses over into the region specifically permitted by the WGC, it still has far to travel before reaching the ``WGC curve'' in Figure 1. Thus there is ample scope for the right side of (\ref{R}) to be usefully reduced.

Evidently, this argument is more of a programme than a calculation. It remains to be substantiated by explicit examples, in which the precise form of the ``WGC curve'' in Figure 1, and consequently the extent of the reduction in the right side of (\ref{R}), can be computed. Nevertheless, we feel that this discussion gives some grounds for optimism.

\section* {\large{\textsf{6. Conclusion}}}
It has always been clear that it will be no easy task to construct a holographic dual for any theory closely resembling QCD; see for example \cite{kn:karch}. However, the AdS/CFT correspondence posits an exact equivalence between string theory in the bulk and a field theory with $N_{\textsf{c}}$ colours. It holds for all values of \emph{all} parameters, including $N_{\textsf{c}}$; so, in principle, holographic duals of QCD-like theories should exist. In practice, one knows how to make use of the duality only when the number of colours is large, but, as is stressed in \cite{kn:mateos} for example, this is usually regarded not as a ``fundamental obstacle'' but rather as a ``technical difficulty''.

Our results here call this claim into question. If Censorship prevents a substantial reduction in $N_{\textsf{c}}$, then, to the extent that classical Censorship for AdS black holes stands firm, we have a fundamental obstacle. We have seen, however, that the WGC promises to convert this obstacle to a (possibly formidable) technical difficulty.

It may be, of course, that field theories with relatively small values of $N_{\textsf{c}}$ do not have duals which are at all similar to AdS$_5$-Reissner-Nordstr\"{o}m spacetimes. Whatever the true dual geometry of such a field theory may be, however, it almost certainly involves an asymptotically AdS$_5$ black hole, which will be subject to its own Censorship restriction; and this too will lead to a bound on the number of colours in the boundary field theory. The problem will then be to understand how this bound can be so low. Again, the WGC will no doubt be crucial here.

\addtocounter{section}{1}
\section*{\large{\textsf{Acknowledgement}}}
The author is grateful to Dr. Soon Wanmei for useful discussions.

\end{document}